# Exploration of Machine Learning Classification Models Used for Behavioral Biometrics Authentication

## SARA G KOKAL

Computer Science department, University of Wisconsin Eau-Claire

Email: kokalsg4814@uwec.edu

## LAURA K PRYOR

Computer Science department, University of Wisconsin Eau-Claire

Email: pryorlk8701@uwec.edu

## RUSHIT DAVE

Computer Science department, University of Wisconsin Eau-Claire

Email: daver@uwec.edu

Mobile devices have been manufactured and enhanced at growing rates in the past decades. While this growth has significantly evolved the capability of these devices, their security has been falling behind. This contrast in development between capability and security of mobile devices is a significant problem with the sensitive information of the public at risk. Continuing the previous work in this field, this study identifies key Machine Learning algorithms currently being used for behavioral biometric mobile authentication schemes and aims to provide a comprehensive review of these algorithms when used with touch dynamics and phone movement. Throughout this paper the benefits, limitations, and recommendations for future work will be discussed.

# 1 INTRODUCTION

In the past decades, mobile touch screen devices have evolved rapidly as demand for rise in technological capability and production has grown. Mobile phones are now capable of storing a variety of information and data relating to emails, photos, video games, and mobile banking applications [1]. Along with this peak in capability, the problem of preserving security in these devices has become apparent. Research into the progression of smartphone security is needed, as the evolution of security of mobile devices must not fall behind. The sensitive information of the public is at risk.

As a response to this concern for security, many methods of mobile device authentication have been produced and analyzed by researchers and smartphone manufacturers. The most common strategies for mobile authentication today are knowledge-based or physiological biometric-based. Knowledge-based uses pins or passwords to grant access to devices [2], and physiological biometrics focus on capturing the unique physical aspects of the user, such as their face or fingerprints [3]. Both these approaches have been met with major security issues. For example, in knowledge-based, the imposter may steal or copy down the user's pin to gain access to their device [2]. Behavioral biometrics, however, focus on how the behaviors of the user can be captured and recorded as data. Behavioral biometrics have been proved to be effective in authentication in multiple systems and devices [4,5], especially when used with Machine Learning algorithms which have also been used in authentication [6,7]. Also, behavioral biometrics can combat the issue of additional hardware costs in physiologic authentication since they primarily use software and machinery that is already in most mobile devices [8]. Other problems regarding knowledge-based and physiological authentication methods are that they are typically one-time authentication schemes. In contrast, behavioral biometrics have the capability of authenticating a user continuously as they use their device [9].

The results of today's research find that behavioral biometrics can provide a secure authentication method if more research is invested into accuracy and tolerance to attacks [2]. The focus of this paper will be on the use of touch and motion dynamics with mobile touch screen devices with various machine learning algorithms. Previous researchers have invested their time in this topic and provided reviews [10], so in according this paper will continue this work by providing a comprehensive review on recent recordings and experimentation related to popular machine learning algorithms with touch and motion dynamics to guide and inform future researchers in their experiments relating to Machine Learning algorithms performance on touch dynamics.

# 2 LITERATURE REVIEW

Before comparing the different Machine Learning algorithms discussed throughout this paper, it is important to provide an overview of these specific algorithms. The papers reviewed feature 4 major ML algorithms: Random Forest (RF), Support Vector Machine (SVM), Gradient Boosting Classifier (GBC), and Multilayer Perceptron (MLP). In papers [11-17], RF was tested on and chosen for further experimentation in their research. Random Forest is an algorithm used for classification and regression problems, able to handle continuous variables for regression, and categorical variables for classification. It uses the ensemble technique, where multiple models are created and then combined to make predictions. For RF, this means that many different individual trees are constructed from each sample, where each tree generates an output. From there, all the outputs are combined and averaged to produce the final output. In a sense, Random Forest is a majority vote. In papers [18-21], SVM is the focus. SVM is used for classification and regression problems, but mainly classification. In SVM, an n-dimensional space is created from the number of features used in the model. From there, the algorithm attempts to find the optimal hyperplane, the decision boundary that helps classify the data points. Support vectors are used to help build the SVM; they are the data points closest to the hyperplane and affect its position and orientation. The optimal hyperplane is one that has the maximum distance between the datapoints of, in the case of the papers discussed, two classification groups. This hyperplane is then used as the decision threshold for new datapoints. The Gradient Boosting classifier is also seen frequently throughout the reviewed papers. GBC uses the ensemble technique, like RF. In the case of GBC, a prediction model is created from several weak prediction models, sometimes in the form of trees. While RF utilizes the bagging ensemble method, GBC uses the boosting method where weak learners are combined into stronger learners so that the final model produces the highest accuracy. MLP is also a highly used classifier. Multilayer Perceptron is a Deep Learning algorithm. Deep Learning algorithms consist of artificial neural networks that attempt to replicate brain structure. MLP is made of input, output, and hidden layers of neurons and is a feedforward algorithm, in that the information moves forward through the nodes in one direction from the input nodes to the output nodes. While the information moves in one direction, the mapping between input and output is not linear.

## 2.1    Random Forest

Researchers ABA Ali et al [11], developed an android application to evaluate what feature and classifier combinations were best suited for continuous swipe-based touch authentication with mobile devices. First, researchers identified all possible swipe-movement-based features before preprocessing them to remove outliers. For the classifier selection phase, features extracted from the raw data were then used with five different classifiers: SVM, Multi-Layer Perceptron (MLP), K-Nearest Neighbor (KNN), Decision Tree (DT), and Random Forest (RF). It was found that RF performed best on all metrics, Frequency Modulation (FM) especially, providing evidence of Random Forest performing well with swipe gesture-based authentication. RF was then used for further experimentation in phase 2 of the system evaluation with feature selection and number of swipes. RF was once again tested with another set of classifiers including Decision Tree (DT) (J48 in WEKA), MLP, Sequential Minimal Optimization (SMO) and KNN. Results found in table 1 found that the model produced with RF and feature selection outperformed the other classifiers as well as comparative studies. The RF model was found to perform best with a threshold value of 40 and 7 to 10 swipes. At threshold 40, no imposters were accepted. Limitations of the study highlighted by the authors included limited touch gestures tested, and the use of a two-class classification system instead of one-class.

Researchers in [12], propose their three-step keystroke and motion-sensor-based authentication scheme. The model first finds the orientation of the mobile device (sitting, walking, and relaxing), then identifies the user's typing position (landscape or portrait). Finally, the user is authenticated using the classification template created based on the orientation and position combination of the user. This model was testing using the KNN and Random Forest (RF) classifiers. Random Forest was found to outperform KNN in initial testing. After the first round of experimentation, Particle Swarm Optimization (PSO) was applied to a feature subset of the model to decrease error rates and number of features used. After optimization, Random Forest outperformed KNN in all metrics, with a best accuracy of 92.58% in relaxing landscape position and a lowest Equal Error Rate (EER) of 2.2% in relaxing landscape position as seen in figure 1. RF also achieved a best False Acceptance Rate (FAR) value of 0.5% in both relaxing landscape and walking landscape positions, demonstrating a 1 in 100 chance of the model failing to recognize an imposter. RF also demonstrated a lowest False Rejection Rate (FRR) of 7.4% in relaxing landscape. Results showed that relaxing and walking positions were most desired for keystroke-based authentication. Ultimately, the study found that determining position before authentication improved metrics, and that their model with RF can efficiently authenticate and satisfy European standards for access control (FAR < 1%). Limitations of the study included a lack of testing against attacks and a limited amount of typing positions tested [12]. Without testing against attacks, researchers cannot be sure that their model is effective in a real-world context where attackers use special methods to get into mobile devices. Testing with limited typing positions makes a model possibly unreliable in a real-world context, as users will type in a multitude of different positions and activities.

Table 1: Accuracy and FM score % comparison with Classifiers [11]

| ML Algorithm | Accuracy | FM |
| --- | --- | --- |
| **RF** | **98.31** | **95.70** |
| J48 | 96.48 | 91.40 |
| MLP | 95.02 | 87.80 |
| SMO | 80.00 | N/A |
| KNN | 97.36 | 93.60 |

In [13], researchers attempted to find out how effective sensor-based information is in authenticating smartphone users. Mostafa et al. designed and proposed their model Behavio2Auth, which utilizes sensor micromovement data while typing to authenticate users implicitly and continuously. The model uses accelerometer data collected from users when typing while sitting or walking. For experimentation, a public Hand Movement, Orientation, and Grasp (HMOG) dataset was used with Random Forest. RF was choses for robustness to noise and outliers as well as its capability of high accuracy. After testing, it was found that performance was better when typing while walking compared to sitting, the results of which can been seen in table 2. While feature selection did not heavily influence accuracy, it was affective in improving the model by reducing features, allowing the classifier to compute faster and more efficiently. Overall, this study demonstrates that RF has decent capability in authenticating based on sensor micromovements. Limitations for this study include a confined range of sensor testing with only accelerometer data analyzed.

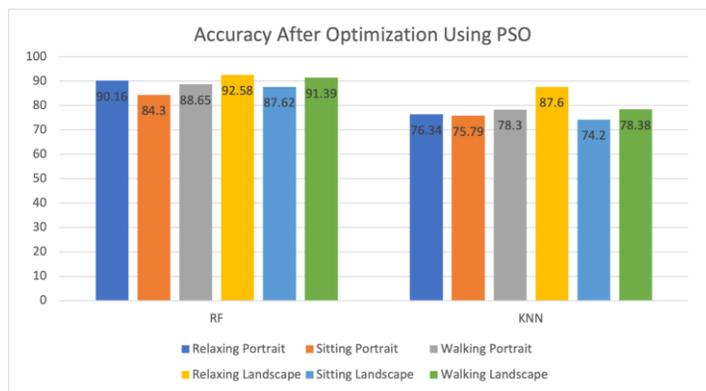

Figure 1: Accuracy after optimization using PSO [12]

Table 2: EER of Random Forest in different scenarios [13]

| Metric | Walking | Sitting |
| --- | --- | --- |
| EER | 0.10 | 0.16 |
| Accuracy | 0.95 | 0.91 |

## 2.2   Support Vector Machine

Researchers in [18], developed their scheme, SwipeVLock, which authenticates users based on swipe dynamics. To unlock their device, users will select their pre-chosen image and then swipe from their pre-chosen location on the image. Their model was tested using DT(J48), SVM, Naive Bayes (NB), and Back Propagation Neural Network (BPNN). In the experimentation phase, 30 participants were separated into 2 groups: Group A performed testing in the lab while Group B set up SwipeVLock in the lab and could keep using the device outside. SVM was found to have the best error results of all metrics as seen in figure 2. Due to these metrics, SVM was chosen as the classifier for the SwipeVLock to its effectiveness and lower error rates with the swipe model. The study also found in the retention phase that the group that was allowed to perform outside the lab (Group B) had better retention 3 days after initial testing with 98% success rate compared Group A's 88%. The limitations of the study included the small sample size of 30 which was then split into smaller groups for experimentation and the lack of testing against various attack types.

In [19], researchers propose their Android pattern lock authentication system, Fine-Grained and Context-Aware Behavioral Biometrics System (FCBBS), using touch and motion dynamics. Their study aimed to combat the problem of using imposter data for training in advance of testing by using One Class-Support Vector Machine (OC-SVM), a one class algorithm not requiring imposter data. While Random Forest was used to detect the context and position of the user, OC-SVM was then used for the identity authentication model. During testing, 4 different patterns (L, Z, S, and T), and 3 contexts (sitting, walking and combination) were used. The system was also tested with and without context detection. Results found that their model performed the best in the sitting context which can been seen in figure 3. When testing for effectiveness with and without context detection using the combined dataset it was found that OC-SVM with context detection produced best results, as seen in figure 3. The researchers also collected data to compare the performance of OC-SVM with other one-class classifiers such as Local Outlier Factor and Isolation Forest. It was found that OC-SVM outperformed the other classifiers in all 4 patterns with a best score of 0.693% EER with T patten. Overall, the study showed the ability of an OC-SVM model to effectively authenticate using touch and motion data. Limitations of the study included a lack of consideration for long term usability as well as the use of limited contexts and patterns for testing.

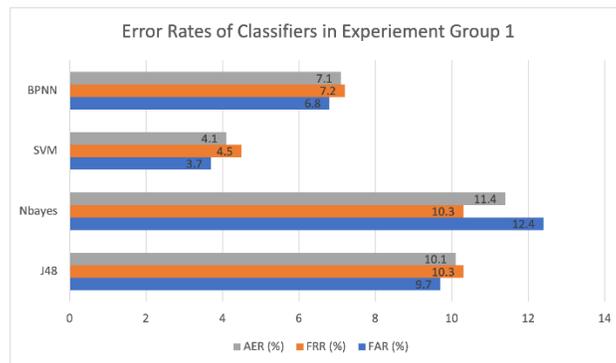

Figure 2: Error Rates of Classifiers in group 1 [18]

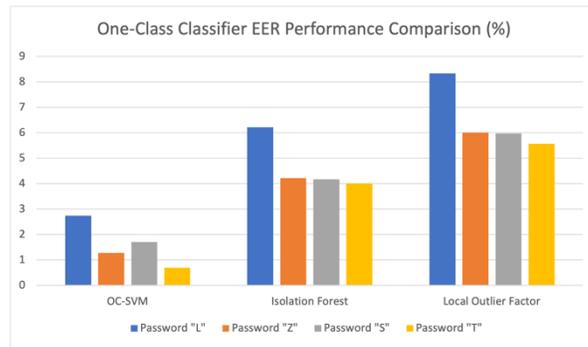

Figure 3: EER Performance of one-class classifier algorithms [19]

In [20], researchers propose their continuous authentication system, DAKOTA, for a mobile banking application. Dakota utilized both motion sensors and scrolling swipe dynamics. The system was tested with classification algorithms such as SVM, KNN, MLP, DT, RF, NB, and Ensemble Learning. Experimentation included evaluation on classifier performance and the effect of posture on model performance. Results found that SVM with both Synthetic Minority Oversampling TEchnique (SMOTE) and random sampling methods had best metrics. SVM with SMOTE had a highest overall accuracy compared with the other algorithms as seen in table 3. SVM was then used for further experimentation, finding that error rates and accuracy were similar with and without combining posture data, and that using a combination of motion and touch data outperformed using individually. Overall, SVM was found to be an effective classifier to use with a touch and motion sensor-based model, with 3.5% EER and 99.8% True Positive Rate (TPR). Limitations of the study included the use of only scroll data for analysis and a lack of consideration for attacks.

## 2.3   Other Algorithms

In [22], researchers trained and tested continuous authentication (CA) models using keystroke data from a public HMOG dataset to compare the effectiveness of different ML algorithms with CA. A total of 8 different machine learning classifiers were tested including ensemble methods such as RF, Extra Trees Classifier (ETC), and GBC as well as KNN, SVM, Classification and Regression Tree (CART), and Naïve-Bayes. Keystroke measurement metrics used for the study were pressingTime, timeReleaseNextPress, and timeBetweenPress. The models were trained using the data from the 8 writing sessions of the HMOG dataset. The results after experimentation showed that ensemble algorithms (RF, ETC, and GBC) performed the best overall with over 70% on most target metrics. Overall, GBC performed the best out of all classifiers in accuracy and target metric recall, with a highest accuracy of 0.71 and substantially high recall values comparatively as seen in figure 4. It is notable that SVM performed the worst with an average accuracy of 0.59. This could suggest that hyperplane classifiers are a poor choice for keystroke based continuous authentication schemes. Similar scores for all metrics were returned by all 7 classifiers, suggesting a lack of bias between false rejections and acceptations. GBC combines multiple weak trees paired with gradient boosting, in contrast to other tree-based classifiers utilizing a forest or a single strong tree, thus, it was argued that the GBC method gives belter results for the keystroke-based CA problem. Limitations of the study included usage of fewer features than current state of art models, and high variability returned by all classifiers. In addition, due to the nature of using data from a public dataset, the authors could not assess imbalances in the data.

Table 3: Classifier Accuracy Comparison with SMOTE sampling [20]

| ML Classifier | Accuracy (%) |
| --- | --- |
| **SVM** | **99.88** |
| MLP | 99.85 |
| KNN | 99.58 |
| RF | 99.49 |
| DT | 99.49 |
| NB | 99.23 |
| MLP-SVM | 99.85 |
| SVM(RBF-POLY) | 99.80 |

Researchers Lamiche et al [23], attempt to address a need for multimodal continuous authentication mechanisms by developing a model using gait patterns and keystroke dynamics. To preserve power consumption, only the accelerometer was used for motion data. Classifiers tested were SVM, RF, Random Tree, Naive-Bayes, and MLP. During the data acquisition phase, 20 participants were asked to perform in 4 different walking scenarios. Due to the nature of this review paper in analyzing ML with touch and motion data, the results from scenario 4, walking while typing, will be focused on. Results found that MLP performed the best in authentication during the walking while typing scenario, as seen in table 4. These results demonstrate MLP efficiency in a multi-modal CA system. When analyzing with keystroke data alone, it was found that MLP was only marginally beat by NB with 0.22% difference in accuracy. Switching to only keystroke data also resulted in a drop of accuracy levels. It is notable that SVM performed the best in identification with 91.50% accuracy. The model of walking and typing was also tested against various attacks such as the zero-effort attack and the minimal effort mimicking attack. Results found that the model demonstrated resistance against these types of attacks with an average FAR of 0.112% under zero effort attack and 0% under minimal effort attack. Ultimately, this study shows potential for gait and keystroke based multimodal CA, as well as MLP proficiency with gait and keystroke CA models. Limitations for this study are a lack of consideration for the effects of disease and injury on the CA model.

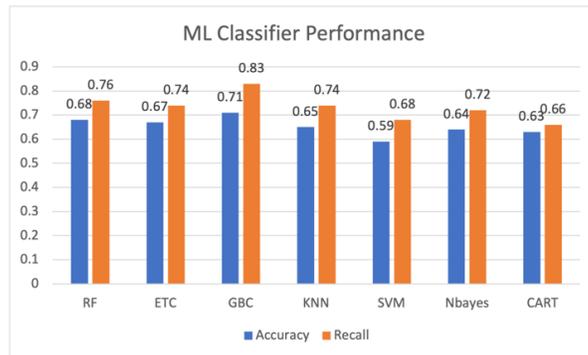

Figure 4: ML classifier performance for CA [22]

Table 4: Classifier performance during Typing while Walking. [23]

| ML Classifier | Accuracy | EER |
| --- | --- | --- |
| SVM | 0.9482 | 0.2688 |
| RF | 0.9713 | 0.0547 |
| RT | 0.9668 | 0.0482 |
| NB | 0.9818 | 0.06 |
| **MLP** | **0.9911** | **0.01** |

## 3 DISCUSSION

Throughout this survey paper, many different algorithms and their current uses in experimental studies related to touch and motion based behavioural biometrics were discussed. As seen in table 5, two algorithms were most frequently used and generally had the best results; SVM and RF. SVM and RF both had high accuracy and promising error rates in the papers reviewed where they were considered the best algorithm. Random Forest was able to perform well in both keystroke and swipe-based dynamics paired with motion sensors. SVM tended to be used in studies focusing on swipe dynamics, of which it performed with high accuracy rates as well. Both algorithms were also able to performance well with continuous authentication-based schemes. It is worth considering that in [22], SVM paired with keystroke-based CA had worse performance in comparison to other classifiers, suggesting that hyperplane classifiers like SVM could be a poor choice for this type of CA scheme. Overall, these two algorithms should be given great consideration for use in touch and motion-based behavioral biometrics schemes. The other algorithms are still worthy of consideration as MLP and GBC were able to outperform both RF and SVM in their respective focus studies. MLP also had decent accuracy compared to RF in [11]. Ultimately, these results show that while SVM and RF currently dominate classification in this subject field, including other classifiers in research testing is important and worthwhile.

## 4 LIMITATIONS & FUTURE WORK

Although the schemes presented in this survey give promising results and notable findings for behavioral biometric authentication, many of the studies had limitations in their research and data collection that could affect the validity of their results. Many of the studies in this survey included a small sample size of limited participants. Smaller sample sizes can affect the data with more variability, leading to bias and running the risk of inaccurately representing the population. In addition, it is important for studies proposing their authentication scheme to test against attacks and while some papers did, the majority did not. Furthermore, it is important to consider that behavior can change over time. Many studies did not evaluate long-term usability. For future work, this survey recommends that researchers testing authentication schemes with behavioral biometrics should focus on extending their sample sizes, engaging in context detection with posture variety, testing against attacks and analyzing long-term usability. These strategies can ensure the scheme's usability in a real-world context.

Table 5: ML Algorithm Summaries

| Papers | Classification Algorithms Tested | Best Classifier | Best Results |
|---|---|---|---|
| [11] | SVM, MLP, KNN, DT, RF | RF | ACC 98.31%, FM 95.70% |
| [12] | KNN, RF | RF | ACC 92.58%, EER 2.2% |
| [13] | RF | RF | ACC 0.95, EER 0.10 |
| [14] | LR, NB, KNN, DT, RF, SVM | RF | ACC 97% |
| [15] | DT, RF, NB, SVM | RF | ACC 95-96% |
| [16] | RF, DT, KNN | RF | ACC 98.8% |
| [17] | BD, DF(RF), SVM, LR | RF | ACC 0.996 |
| [18] | DT(J48), SVM, NB, and BPNN | SVM | 3.7% FAR, 4.5% FRR, 4.1% AER |
| [19] | OC-SVM | SVM | 0.693% EER |
| [20] | SVM, KNN, MLP, DT, RF, NB | SVM | ACC 99.88% |
| [21] | SVM, RF | SVM | ACC 79.88%, FAR 15.84%, and FRR 50% |
| [22] | RF, ETC, GBC, KNN, SVM, CART, NB | GBC | ACC 0.71, Recall 0.83 |
| [23] | SVM, RF, RT, NB, MLP | MLP | ACC 0.9911, EER 0.01 |
| [24] | DT, DVM, KNN, GNB, RF, LR | GNB | EER 2.66% |

## 5 CONCLUSION

To conclude this survey, it was found that Random Forest and SVM outperformed other classifiers and provided high accuracy rates as well as viable error rates. Random Forest performed well in continuous swipe-based authentication as demonstrated by the high accuracy produced in the paper "Smartphone security using swipe behavior-based authentication" [11]. RF also excelled in keystroke authentication paired with motion sensors as shown by the high accuracy produced in "A three-step authentication model for mobile phone user using Keystroke Dynamics" [12], and the low error rates in "Behavio2Auth: Sensor-Based Behavior Biometric Authentication for Smartphones" [13]. SVM specifically excelled in swipe and touch dynamics, with high accuracy in "Dakota: Sensor and touch screen-based continuous authentication on a mobile banking application" [20] and exemplary error rates in "Swipevlock: A supervised unlocking mechanism based on swipe behavior on smartphones" [18] and "Fine-grained and context-aware behavioral biometrics for pattern lock on smartphones" [19]. It is notable that SVM was found to perform poorly in keystroke-based CA with low accuracy as shown in [22] "Comparing machine learning classifiers for continuous authentication on mobile devices by Keystroke Dynamics". Both algorithms were able to perform well in continuous authentication schemes. It is important to also test other classifiers, if possible, as other classifiers such as MLP and GBC were able to outperform RF and SVM in some cases. In "Comparing machine learning classifiers for continuous authentication on mobile devices by Keystroke Dynamics" [22], GBC was able to score highest accuracy and recall compared to both SVM and RF. In [23], MLP was able to score highest accuracy and best error rates compared to SVM and RF. Despite these examples, SVM and RF are ultimately the best choices for touch and motion based mobile device authentication as demonstrated by the results of recent studies analysed in this survey.

## ACKNOWLEDGMENTS

Funding for this project has been provided by the University of Wisconsin-Eau Claire's Blugold Fellowship and University of Wisconsin-Eau Claire Computer Science Department's Karlgaard Scholarship.

## REFERENCES

[1] Nyle Siddiqui, Rushit Dave, and Naeem Seliya. 2021. Continuous User Authentication Using Mouse Dynamics, Machine Learning, and Minecraft. 2021 International Conference on Electrical, Computer and Energy Technologies (ICECET) (2021). DOI:https://doi.org/ https://doi.org/10.48550/arXiv.2110.11080

[2] Nyle Siddiqui, Laura Pryor, and Rushit Dave. 2021. User authentication schemes using machine learning methods—a review. *Algorithms for Intelligent Systems* (August 2021), 703–723. DOI:http://dx.doi.org/10.1007/978-981-16-3246-4_54

[3] Joseph Shelton, Christopher Rice, Jasmeet Singh, John Jenkins, Rushit Dave, Kaushik Roy, and Suryadip Chakraborty. 2018. Palm Print Authentication on a Cloud Platform. 2018 International Conference on Advances in Big Data, Computing and Data Communication Systems (icABCD) (2018). DOI:https://doi.org/10.1109/icabcd.2018.8465479

[4] Janelle Mason, Rushit Dave, Prosenjit Chatterjee, Ieschecia Graham-Allen, Albert Esterline, and Kaushik Roy. 2020. An Investigation of Biometric Authentication in the Healthcare Environment. Array 8, (2020), 100042. DOI:https://doi.org/10.1016/j.array.2020.100042

[5] Joseph M. Ackerson, Rushit Dave, and Naeem Seliya. 2021. Applications of Recurrent Neural Network for Biometric Authentication & Anomaly Detection. Information 12, 7 (2021), 272. DOI:https://doi.org/10.3390/info12070272

[6] Sam Strecker, Willem Van Haaften, and Rushit Dave. 2021. An Analysis of IoT Cyber Security Driven by Machine Learning. Algorithms for Intelligent Systems (2021), 725-753. DOI:https://doi.org/10.1007/978-981-16-3246-4_55

[7] Sam Strecker, Rushit Dave, Nyle Siddiqui, and Naeem Seliya. 2021. A Modern Analysis of Aging Machine Learning Based IoT Cybersecurity Methods. Journal of Computer Sciences and Applications 9, 1 (2021), 16-22. DOI: https://doi.org/10.48550/arXiv.2110.07832

[8] Dylan J. Gunn, Zhipeng Liu, Rushit Dave, Xiaohong Yuan, and Kaushik Roy. 2019. Touch-based Active Cloud Authentication Using Traditional Machine Learning and LSTM on a distributed Tensorflow framework. *International Journal of Computational Intelligence and Applications* 18, 04 (Novermber 2019), 1950022. DOI:http://dx.doi.org/10.1142/s1469026819500226

[9] Jacob Mallet, Laura Pryor, Rushit Dave, Naeem Seliya, Mounika Vanamala, and Evelyn Sowells Boone. 2022. Hold on and Swipe: A Touch-Movement Based Continuous Authentication Schema based on Machine Learning. *2022 Asia Conference on Algorithms, Computing and Machine Learning (CACML)* (January 2022). DOI:https://doi.org/10.48550/arXiv.2201.08564

[10] Laura Pryor, Rushit Dave, Jim Seliya, and Evelyn Sowells Boone. 2021. Machine learning algorithms in User Authentication Schemes. *2021 International Conference on Electrical, Computer and Energy Technologies (ICECET)* (December 2021). DOI:https://doi.org/10.48550/arXiv.2201.08564

[11] Adnan Bin Amanat Ali et al. 2021. Smartphone security using swipe behavior-based authentication. *Intelligent Automation & Soft Computing* 29, 2 (June 2021), 571–585. DOI:http://dx.doi.org/10.32604/iasc.2021.015913

[12] Baljit Singh Saini et al. 2020. A three-step authentication model for mobile phone user using Keystroke Dynamics. *IEEE Access* 8 (July 2020), 125909–125922. DOI:http://dx.doi.org/10.1109/access.2020.3008019

[13] Hebatollah Mostafa, Mohammad El-Ramly, Abeer Mohamed Elkorany, and Hassan Shaban. 2019. Behavio2Auth: Sensor-Based Behavior Biometric Authentication for Smartphones. *ArabWIC 2019: Proceedings of the ArabWIC 6th Annual International Conference Research Track* (March 2019), 1–6. DOI:http://dx.doi.org/10.1145/3333165.3333176

[14] Kanlun Wang, Lina Zhou, Dongsong Zhang, Zhihui Liu and Jaewan Lim. 2020. What is More Important for Touch Dynamics based Mobile User Authentication? PACIS 2020 Proceedings (June 2020).

[15] Dongsong Zhang, Lina Zhou, and Sailakshmi Pisupati. 2019. Tracing One's Touches: Continuous Mobile User Authentication Based on Touch Dynamics. *Twenty-fifth Americas Conference on Information Systems, Cancun, 2019* (July 2019).

[16] Akriti Verma, Valeg Moghaddam, and Adnan Anwar. 2021. Data-driven behavioural biometrics for continuous and adaptive user verification using Smartphone and Smartwatch. *CoRR* (2021). DOI:https://doi.org/10.48550/arXiv.2110.03149

[17] Nurhak Karakaya, Gülfem Işıklar Alptekin, and Özlem Durmaz İncel. 2019. Using behavioral biometric sensors of mobile phones for user authentication. *Procedia Computer Science* 159 (January 2019), 475–484. DOI:http://dx.doi.org/10.1016/j.procs.2019.09.202

[18] Wenjuan Li, Jiao Tan, Weizhi Meng, Yu Wang, and Jing Li. 2019. Swipevlock: A supervised unlocking mechanism based on swipe behavior on smartphones. *Machine Learning for Cyber Security* 11806 (September 2019), 140–153. DOI:http://dx.doi.org/10.1007/978-3-030-30619-9_11

[19] Dai Shi et al. 2021. Fine-grained and context-aware behavioral biometrics for pattern lock on smartphones. *Proceedings of the ACM on Interactive, Mobile, Wearable and Ubiquitous Technologies* 5, 1 (March 2021), 1–30. DOI:http://dx.doi.org/10.1145/3448080

[20] Ozlem Durmaz Incel et al. 2021. Dakota: Sensor and touch screen-based continuous authentication on a mobile banking

application. *IEEE Access* 9 (March 2021), 38943–38960. DOI:http://dx.doi.org/10.1109/access.2021.3063424

[21]  Mohamed Azard Rilvan, Jedrik Chao, and Md Shafaeat Hossain. 2020. Capacitive swipe gesture based smartphone user authentication and Identification. *2020 IEEE Conference on Cognitive and Computational Aspects of Situation Management (CogSIMA)* (October 2020). DOI:http://dx.doi.org/10.1109/cogsima49017.2020.9215998

[22] Luis de-Marcos, José-Javier Martínez-Herráiz, Javier Junquera-Sánchez, Carlos Cilleruelo, and Carmen Pages-Arévalo. 2021. Comparing machine learning classifiers for continuous authentication on mobile devices by Keystroke Dynamics. *Electronics* 10, 14 (2021), 1622. DOI:http://dx.doi.org/10.3390/electronics10141622

[23] Imane Lamiche, Guo Bin, Yao Jing, Zhiwen Yu, and Abdenour Hadid. 2018. A continuous smartphone authentication method based on gait patterns and Keystroke Dynamics. *Journal of Ambient Intelligence and Humanized Computing* 10, 11 (November 2018), 4417–4430. DOI:http://dx.doi.org/10.1007/s12652-018-1123-6

[24] Yeeun Ku, Leo Hyun Park, Sooyeon Shin, and Taekyoung Kwon. 2019. Draw it as shown: Behavioral Pattern Lock for Mobile user authentication. *IEEE Access* 7 (May 2019), 69363–69378. DOI:http://dx.doi.org/10.1109/access.2019.2918647